\documentclass[a4paper]{jpconf}
\usepackage{graphicx}
\begin{document}
\title{MiniBooNE}
\author{M~Sorel\footnote[1]{Present address: Instituto de Fisica Corpuscular
 (IFIC), CSIC and Universidad de Valencia, Spain.} 
}
\address{Department of Physics, Columbia University, New York, NY 10027, USA} 
\ead{sorel@fnal.gov}
\begin{abstract}
The physics motivations, design, and status of the Booster Neutrino Experiment at Fermilab, MiniBooNE, are briefly discussed. Particular emphasis is given on the ongoing preparatory work that is needed for the MiniBooNE muon neutrino to electron neutrino oscillation appearance search. This search aims to confirm or refute in a definitive and independent way the evidence for neutrino oscillations reported by the LSND experiment.
\end{abstract}
\section{Introduction}
The main goal of the MiniBooNE experiment is to unambiguously confirm or refute the evidence for $\bar{\nu}_{\mu}\to\bar{\nu}_e$ oscillations reported by the LSND experiment \cite{Aguilar:2001ty}. The outcome of this cross-check is important since the LSND result is incompatible with the simplest three-neutrino mixing paradigm  based upon the robust evidence for solar and atmospheric oscillations, and its confirmation by MiniBooNE would have a profound impact on our understanding of neutrinos. \\
\indent The MiniBooNE neutrino beam is a high-intensity, conventional neutrino beam produced via the decay of mesons and muons in a 50 m long decay region following the target hall, where meson production and focusing occurs. Mesons are produced in the interactions of 8 GeV protons from the Fermilab Booster accelerator in a thick beryllium target, and then focused by a magnetic focusing horn surrounding the target. The decay region is instrumented with a spectrometer and range stack detector ("little muon counter", LMC) to measure muons from kaon decays. As of November, 2005, about $6.6\cdot 10^{20}$ protons have been sent to the MiniBooNE target. More details on the Fermilab Booster neutrino beam can be found in Ref.~\cite{Sorel:1900sg}. \\
\indent The MiniBooNE neutrino detector consists of a carbon steel spherical tank of 6.1 m radius and filled with approximately 800 tons of undoped mineral oil. The center of the detector is located at a distance of 541 m from the neutrino production target, below a dirt overburden of about 3 m. Neutrino interactions in the oil are observed by detecting the Cherenkov and scintillation light produced by neutrino-induced charged tracks travelling in the detector medium. An array of 1,280 photomultiplier tubes (PMTs), located at a radius of 5.75 m and oriented toward the center of the tank, is used to record the number and arrival time of the photons produced in the fiducial volume of the detector. The PMTs provide a uniform, 10\% coverage of the whole detector spherical inner surface. The optically isolated, outer detector region is used to reject cosmic-ray induced activity in the tank. The photoelectron charge and time of the PMT signals are continuously digitized and recorded for every proton beam spill. A laser system, a cosmic ray muon hodoscope, and seven scintillation cubes located inside the detector, are used to calibrate various aspects of the MiniBooNE detector response. As of November, 2005, about $7.0\cdot 10^5$ candidate neutrino interactions have been identified in MiniBooNE. More details on the MiniBooNE detector can be found in Ref.~\cite{MiniBooNETDR}. \\
\indent The MiniBooNE experiment is probing the oscillation parameter space indicated by LSND via a $\nu_{\mu}\to\nu_e$ search, and possibly by a future $\bar{\nu}_{\mu}\to\bar{\nu}_e$ search. It is estimated that the large sample of neutrino interactions detected at MiniBooNE will allow coverage of the full LSND 90\% C.L. allowed region at 4 $\sigma$ significance \cite{runplan}. The detailed sensitivity study of Ref.~\cite{runplan} includes the expected neutrino statistics for $10^{21}$ protons on target, background level estimates from particle misidentification and intrinsic $\nu_e$'s in the beam, systematic uncertainties associated with those backgrounds, efficiency in reconstructing and identifying the oscillation signal, and knowledge of the neutrino energy distribution shapes of the various background components and of the eventual oscillation signal for any given $\Delta m^2$ neutrino mass splitting hypothesis. In the case of a confirmation of the LSND signal, MiniBooNE will also provide a rough determination of the mass and mixing parameters responsible for neutrino oscillations, as also shown in Ref.~\cite{runplan}.
%
%
\section{Status of the MiniBooNE $\nu_{\mu}\to\nu_e$ analysis}
The MiniBooNE collaboration is committed to performing a "closed box" analysis for the $\nu_e$ appearance search. For this procedure, the small fraction of data events that could potentially contain electron neutrino interactions in the oscillation signal energy region are "censored" and not available for a full, detailed analysis. The plan is to complete the full development of the analysis with Monte Carlo simulated events, "open" neutrino data, calibration data, and data external to the MiniBooNE experiment before considering the MiniBooNE "closed box" events. The goal is to fully substantiate, and where possible, improve on, the physics assumptions that are implicit in the oscillation sensitivity studies of Ref.~\cite{runplan}. In the following, a flavour of the ongoing preparatory work on Monte Carlo tuning and algorithms optimization that is necessary before "opening the box" is given, and illustrated with some specific examples. \\
\indent First, accurate knowledge of the muon and electron neutrino fluxes at the MiniBooNE detector as a function of energy is necessary for the $\nu_{\mu}\to\nu_e$ analysis. The tool used for neutrino flux predictions is a GEANT4 \cite{Agostinelli:2002hh} description of the beamline, simulating the interactions, focusing and decays of baryons, mesons, and muons in the target hall and decay regions. Pion and kaon production data on beryllium are the most important external physics inputs to the simulation. Pion production data is being tuned on recent data from the E910 experiment at BNL \cite{e910} and the HARP experiment at CERN \cite{harp}. Moreover, the overall muon neutrino flux normalization, and the electron neutrino flux component due to muon decays, are being tuned on MiniBooNE muon neutrino, charged-current, quasi-elastic (CCQE) data. Finally, the electron neutrino flux component due to kaon decays is being tuned on MiniBooNE LMC monitoring data and high-energy CCQE data. In the longer term, the HARP experiment will also provide important inputs on kaon production cross-sections in proton-beryllium interactions. \\
\indent The second key ingredient in the oscillation search is the accurate modelling of neutrino interactions in the MiniBooNE detector. The tool used for this purpose is the NUANCE cross-section generator \cite{Casper:2002sd}, simulating all relevant neutrino interaction channels in the $\sim 1$ GeV energy region, including a detailed treatment of carbon nuclear effects. Several external constraints are used to tune various aspects of the simulation: free nucleon cross-sections from neutrino data, nuclear model from electron data, and final state interactions from pion/proton scattering data. Given that one of the major backgrounds to the oscillation search are neutral-current, single $\pi^0$ production neutrino interactions where the $\pi^0\to\gamma\gamma$ decay is identified as a single electron, particular attention is being given to this neutrino interaction type. The MiniBooNE detector can directly measure the separate amounts of resonant and coherent cross-section contributions to this interaction channel and the $\pi^0$ production kinematics \cite{Raaf:2005up}, therefore allowing to constrain the amount of the dominant misidentification background to the $\nu_{\mu}\to\nu_e$ search. \\
\indent An accurate description of the detector response is obtained via a hit-level GEANT3 \cite{geant3} simulation, characterized by extended user control over the simulation of the detector optical model, that is of the light production and transmission mechanisms in the MiniBooNE mineral oil, and of the photomultiplier charge and time response. Several tabletop measurements have been carried out to constrain critical optical model parameters, such as time-resolved fluorescence, Rayleigh and Raman scattering, and light attenuation as a function of wavelength \cite{Brown:2004uy}. In addition, the overall detector response tuning is accomplished using MiniBooNE calibration data. For example, the simulated total PMT charge distribution is cross-checked with a clean sample of Michel electrons from muon decays at rest, while the simulated angular distribution of light is validated via a sample of cosmic ray muon tracks of known direction. \\
\indent Simulated and actual neutrino interactions are reconstructed based on the measured PMT charge and time pattern. Various algorithms are used to optimize reconstruction and particle identification (PID), in order to achieve good reconstruction accuracy, robustness, and  efficiency of PID separation \cite{Yang:2005nz}. As an example of reconstruction accuracy, studies of simulated neutrino CCQE interactions show that about a 10\% neutrino energy resolution can be achieved. Particle identification robustness is being accomplished by choosing PID variables that are largely insensitive to small variations in the parameters describing the detector optical model and neutrino cross-sections, and variables that exhibit a good match between simulated and actual data from a number of calibration and open neutrino data samples. These same data samples are also used to validate the efficiency of PID separation predicted by Monte Carlo simulations.
%
%
\section{Conclusions}
The basic ingredients for the MiniBooNE $\nu_{\mu}\to\nu_e$ oscillation analysis are already in place. The focus is now on the fine-tuning of the Monte Carlo simulation and reconstruction/PID tools that are used in the analysis, on evaluating and reducing the systematic uncertainties, and on devising consistency checks and increase the analysis robustness. The MiniBooNE $\nu_{\mu}\to\nu_e$ search is a "closed box" analysis. The oscillation results will be published shortly after the box will be opened. The current estimate for opening the box is sometime in 2006. Based on our studies, we expect that already the first MiniBooNE oscillation results will address the LSND evidence for oscillations in a definitive way.
%
%
\ack
The author would like to thank the Organizing Committee for the invitation and for the stimulating atmosphere of the TAUP 2005 conference. The MiniBooNE collaboration gratefully acknowledges support from various grants and contracts from the US Department of Energy and the US National Science Foundation. The author is supported by a Marie Curie Intra-European Fellowship within the 6th European Community Framework Program.
%
%
\section*{References}
\end{document}